\documentclass[aps,prl,showpacs,twocolumn,superscriptaddress]{revtex4}
\usepackage{graphicx}
\usepackage{amsmath,mathrsfs}
\begin{document}

\title{Kondo Effect and Josephson Current through a Quantum Dot between
  Two Superconductors}%

\author{Mahn-Soo Choi}%
\affiliation{Department of Physics, Korea University, Seoul 136-701,
  Korea}%

\author{Minchul Lee}%
\affiliation{Department of Physics, Korea University, Seoul 136-701,
  Korea}%

\author{Kicheon Kang}%
\affiliation{Department of Physics and Institute for Condensed Matter
  Theory, Chonnam National University, Kwangju
  500-757, Korea}%

\author{W. Belzig}%
\affiliation{Department of Physics and Astronomy, University of Basel,
  Klingelbergstrasse 82, 4056 Basel, Switzerland}%

\date{\today}

\begin{abstract}
We investigate the supercurrent through a quantum dot for the whole
range of couplings using the numerical renormalization group method.  We
find that the Josephson current switches abruptly from a $\pi$- to a
$0$-phase as the coupling increases. At intermediate couplings the total
spin in the ground state depends on the phase difference between the two
superconductors. Our numerical results can explain the crossover in the
conductance observed experimentally by Buitelaar \textit{et al.} [Phys.
Rev. Lett.  \textbf{89}, 256 801 (2002)].
\end{abstract}

\pacs{74.50.+r, 72.15.Qm, 75.20.Hr}
%% 74.50.+r Tunneling phenomena; point contacts, weak links, Josephson
%% effects (for SQUIDs, see 85.25.Dq; for Josephson devices, see
%% 85.25.Cp; for Josephson junction arrays, see 74.81.Fa)
%%
%% 72.15.Qm Scattering mechanisms and Kondo effect (see also 75.20.Hr
%% Local moments in compounds and alloys; Kondo effect, valence
%% fluctuations, heavy fermions in magnetic properties and materials)
%%
%% 75.20.Hr Local moment in compounds and alloys; Kondo effect, valence
%% fluctuations, heavy fermions (see also 72.15.Qm Scattering mechanisms
%% and Kondo effect in electronic conduction of metals and alloys)

\maketitle

%%%%
\let\up=\uparrow \let\down=\downarrow \let\eps=\epsilon
\let\veps=\varepsilon
\newcommand\QD{\mathrm{QD}} \newcommand\BCS{\mathrm{BCS}}
\newcommand\bfk{\mathbf{k}} \newcommand\varH{\mathscr{H}}
\newcommand\varS{\mathscr{S}} \newcommand\tilV{\widetilde{V}}
\newcommand\vtilH{\widetilde\varH}
\newcommand\avg[1]{\langle{\textstyle#1}\rangle}

%%%%
\paragraph{Introduction.}
The Kondo effect and superconductivity are two of the most extensively
studied phenomena in condensed matter physics ever since the pioneering
works by Kondo~\cite{Kondo64a} and by Bardeen, Cooper and Schrieffer
\cite{Bardeen57a},
%% ,Bardeen57b},
respectively.  When a localized spin is coupled to superconducting
electrons, the two effects are intermingled and even richer physics will
emerge.  The physically interesting questions are: Would the Kondo
effect survive, overcoming the spin-singlet pairing of electrons in
superconductors (SCs) and the superconducting gap at the Fermi level?  If it
does, how would such a strongly correlated state affect the transport,
especially the Josephson current, between two superconductors?

The Josephson effect through a strongly interacting region with a
localized spin was discussed long before by Shiba and
Soda~\cite{Shiba69a} and Glazman and Matveev~\cite{Glazman89a} and
further elucidated by Spivak and Kivelson~\cite{Spivak91a}.  The large
on-site interaction only allows the electrons in a Cooper pair to tunnel
one by one via virtual processes in which the spin ordering of the
Cooper pair is reversed, leading to a negative Josephson coupling (i.e.,
a $\pi$-junction).  This argument, however, is based on a perturbative
idea and holds true only for sufficiently weak tunneling.  It was
suggested~\cite{Glazman89a} that as the tunneling increases, the Kondo
effect produces a collective resonance at the Fermi level.  As a result,
the Josephson current is enhanced by the Coulomb repulsion.  Moreover,
the Josephson coupling is expected to be positive (i.e., a $0$-junction)
since the localized spin is screened due to the Kondo effect.
Based on this, Glazman and Matveev~\cite{Glazman89a} assumed a strong
coupling fixed point and derived the Josephson current as a function of
phase difference.  Recently, several approximation methods have been
used to investigate the transition from the $0$- to $\pi$-junction as a
function of the tunneling
strength~\cite{Rozhkov99a,Clerk00b,Rozhkov00a,Vecino03a}:
A modified Hartree-Fock
approximation~\cite{Rozhkov99a}, a non-crossing approximation
(NCA)~\cite{Clerk00b}, and a variational method~\cite{Rozhkov00a}
predict a $0$-$\pi$ transition, whereas the slave-boson mean-field
theory~\cite{Rozhkov00a} always favors the Kondo effect.

In this work, we use a numerical renormalization group (NRG) method to
investigate thoroughly the $0$-$\pi$ transition as well as to examine
the argument above suggested by Glazman and Matveev~\cite{Glazman89a}.
Based on the NRG method, we calculate quantitatively the local
properties (i.e., the pairing correlation and the single-particle
excitation spectrum) of the quantum dot (QD), the total spin in the
ground-state wave function, and the Josephson current as a function of
phase difference.
Finally, we show that our numerical results can explain the
experimentally observed crossover of the conductance in SC-carbon
nanotube-SC junctions \cite{Buitelaar02a}.

%%%%
\paragraph{Model.}
The system consists of a QD with an odd number of electrons coupled to
two superconducting leads ($L$ and $R$).
The study of Kondo effect in such a mesoscopic system has recently
attracted much interest due to its tunability.  As already demonstrated
experimentally with normal leads~\cite{Goldhaber-Gordon98a}, it allows
for various tests of Kondo physics, which are difficult in bulk solids.
The two leads are regarded to be standard $s$-wave superconductors (SCs)
and described by the BCS Hamiltonian
\begin{multline}
%%\begin{equation}
\label{kondo-sds::eq:HBCS}
\varH_\BCS = \sum_{\ell=L,R}\sum_{\bfk,\sigma} \eps_{\ell,\bfk}
c_{\ell,\bfk,\sigma}^\dag c_{\ell,\bfk\sigma} \\\mbox{} %
- \sum_\ell\sum_\bfk\left( \Delta_\ell e^{+i\phi_\ell}
  c_{\ell,\bfk,\up}^\dag c_{\ell,-\bfk,\down}^\dag + h.c. \right) \,,
%%\end{equation}
\end{multline}
where $c_{\ell,\bfk,\sigma}^\dag$ ($c_{\ell,\bfk,\sigma}$) creates
(destroys) an electron with energy $\eps_{\ell,\bfk}$, momentum
$\hbar\bfk$, and spin $\sigma$ on the lead $\ell$.  $\Delta_\ell$ is the
superconducting gap and $\phi_\ell$ is the phase of the superconducting
order parameter.
The QD is described by an Anderson-type impurity model
\begin{equation}
\label{kondo-sds::eq:HD}
\varH_\QD = \sum_\sigma \eps_d d_\sigma^\dag d_\sigma
+ U\, d_\up^\dag d_\up\, d_\down^\dag d_\down \,,
\end{equation}
which is widely adopted for sufficiently small quantum dots.  In
Eq.~(\ref{kondo-sds::eq:HD}) $d_\sigma^\dag$ and $d_\sigma$ are electron
creation and annihilation operators on the QD.  The level position
$\eps_d$, measured from the Fermi energy $E_F$ of the two leads
(throughout the paper every energy is measured from $E_F$), can be tuned
by an external gate voltage.  The interaction $U$ is order of charging
energy $e^2/2C$ ($C$ is the capacitance of the QD).
The coupling between the QD and the SCs is described by the tunneling
Hamiltonian
\begin{equation}
\label{kondo-sds::eq:HV}
\varH_V = \sum_\ell\sum_{\bfk,\sigma} V_\ell
\left(d_\sigma^\dag c_{\ell,\bfk,\sigma} + h.c.\right) \,.
\end{equation}
Putting all together the Hamiltonian for the whole system is given by
\begin{math}
\varH = \varH_\QD + \varH_\BCS + \varH_V.
\end{math}

We take a few simplifications to make clearer the physical
interpretation of the results below.  The two SCs are assumed to be
identical ($\eps_{L,\bfk}=\eps_{R,\bfk}=\eps_\bfk$ and
$\Delta_L=\Delta_R=\Delta$) except for a finite phase difference
$\phi=\phi_L-\phi_R$; without loss of generality we put
$\phi_L=-\phi_R=\phi/2$.  In the normal state, the conduction bands on
the leads are symmetric with a flat density of states $N_0$ and the
width $D$ above and below the Fermi energy.  We also put $\eps_d=-U/2$
in $\varH_\QD$, Eq.~(\ref{kondo-sds::eq:HD}); it has been checked that
an asymmetric model ($\eps_d\neq -U/2$) gives the qualitatively same
results for physical quantities of our concern.  We only consider the
symmetric junction, $V_L=V_R=V$.  The coupling to the leads is well
characterized by the single parameter $\Gamma = 2\pi N_0 V^2$.  Below we
will distinguish the strong ($T_K\gg\Delta$) and the weak
($T_K\ll\Delta$) coupling limits by the ratio between the
superconducting gap $\Delta$ and the \emph{normal-state} Kondo
temperature $T_K$ ($k_B=1$) given by~\cite{Haldane78a}
\begin{equation}
\label{TwoLeads::eq:TK 2}
T_K = \Gamma\,
\sqrt\frac{U}{2\Gamma}\,
\exp\left[\pi\, \frac{\eps_d}{2\Gamma}\,
  \left(1+\frac{\eps_d}{U}\right)
\right] \,.
\end{equation}

Following the standard NRG procedures\cite{Wilson75a,Krishna-murthy80a}
extended to superconducting leads \cite{Yoshioka00a}, we evaluate the
various physical quantities from the recursion relation
\begin{multline}
\label{kondo-sds::eq:HNRG}
\vtilH_{N+1} = \sqrt\Lambda\, \vtilH_N + \xi_N\sum_{\mu,\sigma}
\left(f_{\mu,N,\sigma}^\dag f_{\mu,N+1,\sigma} + h.c.\right) \\\mbox{} %
- \Lambda^{N/2}\sum_\mu\widetilde\Delta_\mu \left(f_{\mu,N+1,\up}^\dag
  f_{\mu,N+1,\down}^\dag + h.c.\right)
\end{multline}
with the initial Hamiltonian given by
\begin{multline}
%%\begin{equation}
\label{kondo-sds::eq:H0}
\vtilH_0 = \frac{1}{\sqrt\Lambda} \Biggl[ \vtilH_\QD +
\sum_{\mu=e,o}\sum_\sigma\tilV_\mu \left( d_\sigma^\dag f_{\mu,0,\sigma}
  + h.c. \right) \\\mbox{} %
- \sum_{\mu}\widetilde\Delta_\mu
\left(f_{\mu,0,\up}^\dag f_{\mu,0,\down}^\dag + h.c.\right) \Biggr] \,.
%%\end{equation}
\end{multline}
Here the fermion operators $f_{\mu,N,\sigma}$ have been introduced as a
result of the logarithmic discretization and the accompanying canonical
transformation, $\Lambda$ is the logarithmic discretization parameter
(we choose $\Lambda=2$), $\xi_N\sim 1$~\cite{Wilson75a}, and
\begin{eqnarray}
\vtilH_\QD \equiv \zeta
\frac{\varH_\QD}{D} &, &
\widetilde\Delta_\mu \equiv \zeta \frac{\Delta_\mu}{D} \,,\\\nonumber
\tilV_e \equiv \zeta
\sqrt{\frac{2\Gamma}{\pi D}}\cos(\phi/4) &,&
\tilV_o \equiv -\zeta
\sqrt{\frac{2\Gamma}{\pi D}}\sin(\phi/4)\,,
\end{eqnarray}
with $\zeta=\frac{2}{1+1/\Lambda}$.  The Hamiltonians $\vtilH_N$ in
Eq.~(\ref{kondo-sds::eq:HNRG}) have been rescaled for numerical
accuracy.  The original Hamiltonian is recovered by
\begin{math}
\varH/D = \lim_{N\to\infty}\vtilH_N/\varS_N
\end{math}
with
\begin{math}
\varS_N \equiv \zeta\Lambda^{(N-1)/2} \,.
\end{math}

%%%%
\begin{figure}
\centering%
\includegraphics*[width=60mm]{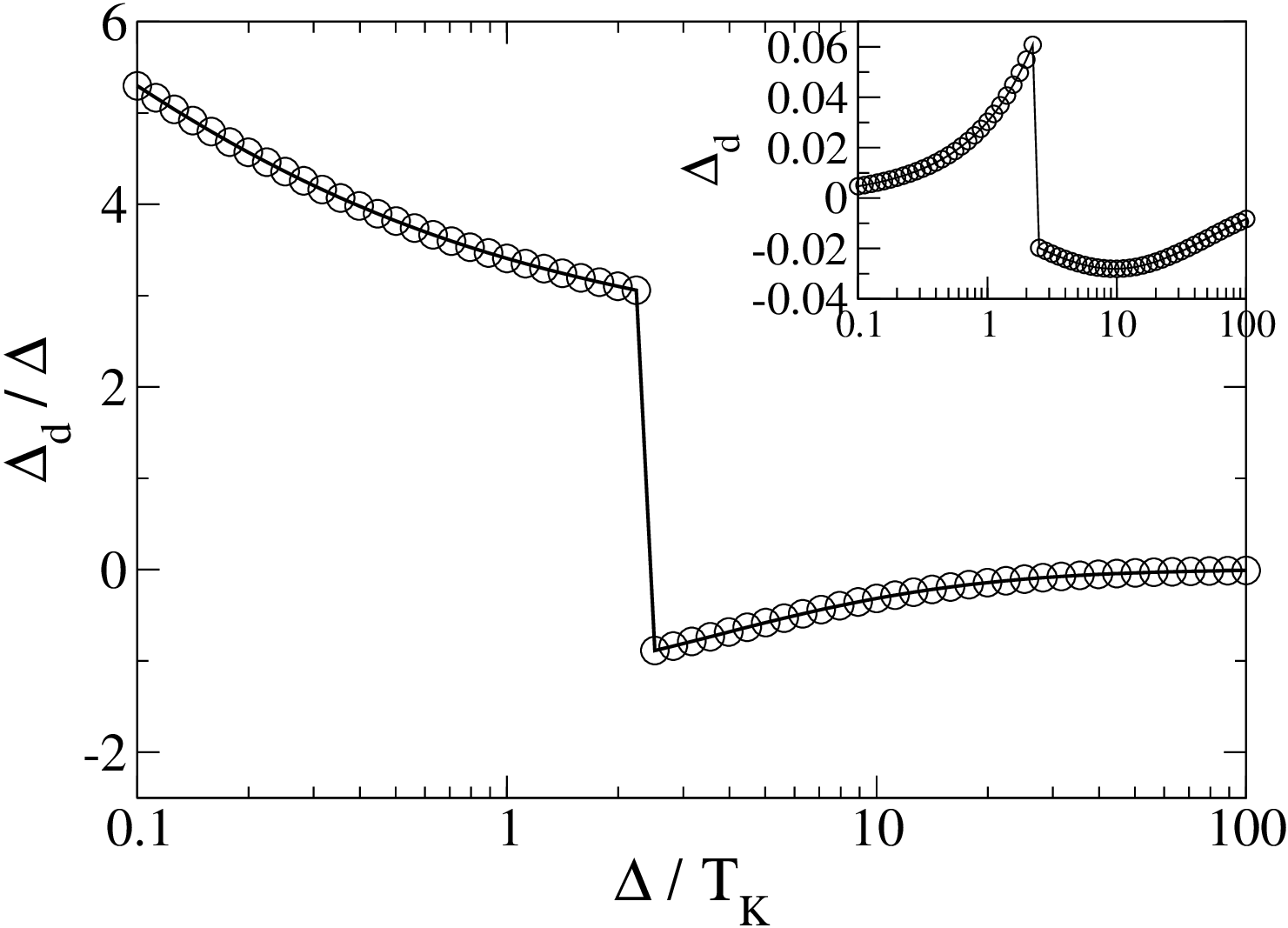}%
\caption{The pairing correlation on the quantum dot,
  $\Delta_d\equiv\avg{d_\up^\dag d_\down^\dag}$, as a function of
  $\Delta/T_K$. Inset: plot of bare (not normalized) values of
  $\Delta_d$.  We have chosen $\eps_d=-U/2=-0.1D$ and $\Gamma=0.04D$.}
\label{kondo-sds::fig:1}
\end{figure}

\paragraph{Proximity effect.}
To see how superconductivity on the leads affects the interacting QD in
the strong and weak coupling limits, we first examine the local
properties on the QD with zero phase difference
($\phi=0$)~\cite{endnote:1}.  Figure~\ref{kondo-sds::fig:1} shows the
local pair correlation
\begin{math}
\Delta_d \equiv \avg{d_\up^\dag d_\down^\dag}
\end{math}
as a function of $\Delta/T_K$.  As expected, the local pair correlation
$\Delta_d$ vanishes with $\Delta$, and gets smaller (even vanishes when
$U\to\infty$) as $\Delta\to\infty$; see Fig.~\ref{kondo-sds::fig:1}
(inset).  An interesting aspect of $\Delta_d$ is the sign change at
$\Delta=\Delta_c\simeq 2.4T_K$, which suggests that the physical
properties are different in the strong ($T_K\gg\Delta$) and the weak
($T_K\ll\Delta$) coupling limits.  Indeed we see (from the NRG
calculation) that the ground-state wave function of the whole system is
of spin singlet (the localized spin is screened out) for
$\Delta<\Delta_c$ and of spin doublet (the SCs form Cooper pairs
separately and the localized spin is left unscreened) for
$\Delta>\Delta_c$.  The negative sign in $\Delta_d$ in the weak coupling
limit can be explained by a simple second-order perturbation theory,
while the positive on in the strong-coupling limit is expected when
there is a resonance channel for Cooper-pair
tunneling~\cite{Beenakker91a}.  Therefore, it seems quite plausible to
argue that in the strong coupling limit the Kondo resonance develops
even in the presence of the superconducting gap in the conduction band
and the proximity effect arises through the resonance; see also the
discussion of the Josephson current below.
Putting it another way, the local moment of spin 1/2 induces a negative
$\Delta_d$ for weak couplings, but as the coupling increases it is
screened and a positive $\Delta_d$ is recovered.

\begin{figure}
\centering
\includegraphics*[width=65mm]{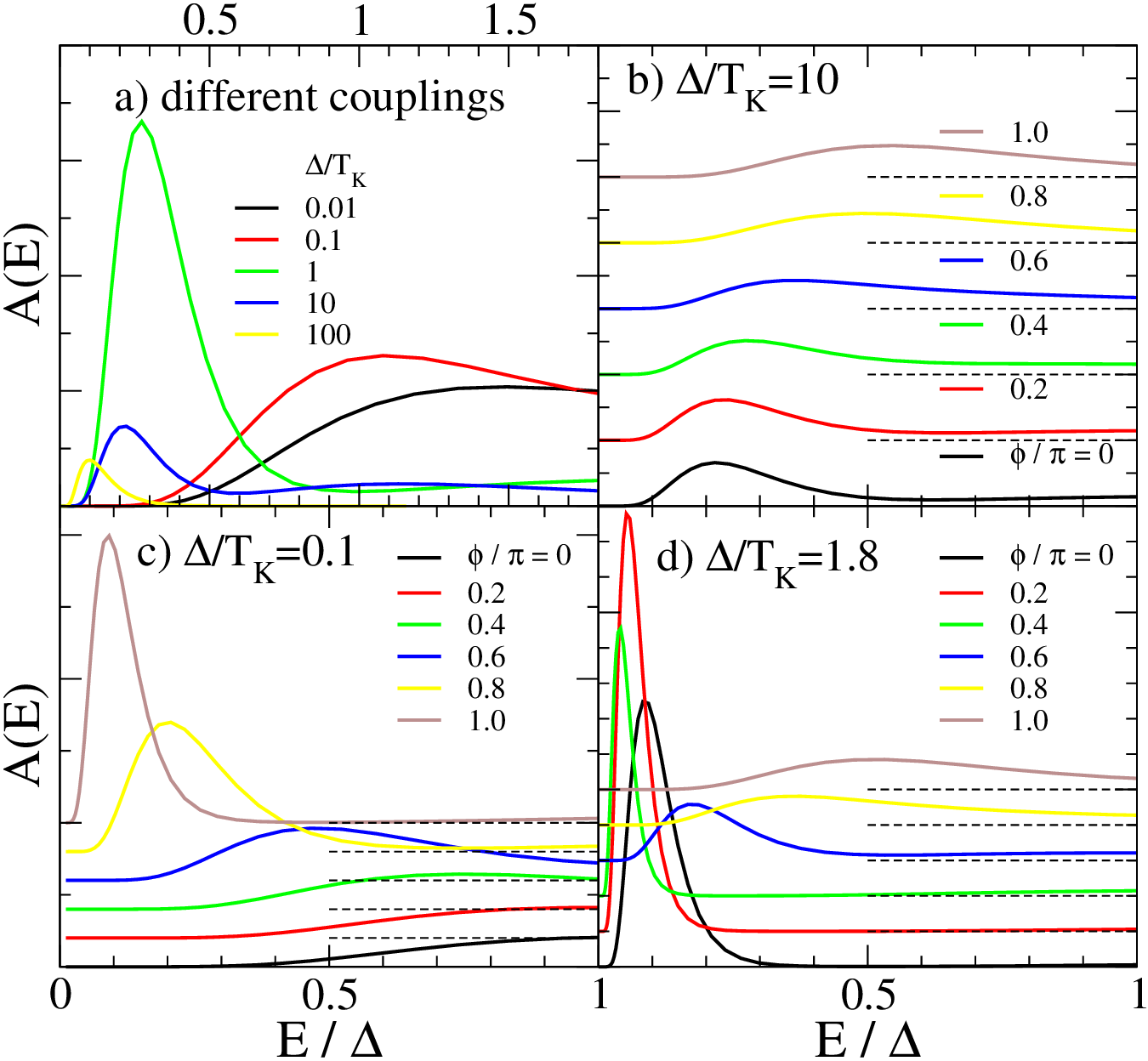}%
\caption{The single-particle excitation spectrum on the quantum dot.
  $\eps_d=-U/2=-0.1D$ and $\Gamma=0.04D$ ($T_K=0.0089D$).}
\label{kondo-sds::fig:2}
\end{figure}

This interpretation is further supported by the single-particle
excitation spectra $A_d(E)$ on the QD, as shown in
Fig.~\ref{kondo-sds::fig:2} for different values of $\Delta/T_K$.  In
Fig.~\ref{kondo-sds::fig:2} (a) $A(E)$ for zero phase
difference~\cite{Sakai93a} is shown and we observe a qualitative change
of the spectrum when $\Delta$ becomes smaller than $T_K$. A localized
state \textit{below} the superconducting gap appears for $\Delta\gtrsim
T_K$, whereas the spectrum has a gap of the order of $\Delta$ in the
other limit.  The other panels in Fig.~\ref{kondo-sds::fig:2} show the
phase-dependent density of states in the sub-gap regime. We clearly
observe a phase-dependent formation of an Andreev bound state. For
$\Delta/T_K=0.1$ the Andreev state emerges from the gap with increasing
phase and reaches the smallest energy for $\phi=\pi$, which is
reminiscent of a usual superconducting junction. In weak coupling limit,
$\Delta=10 T_K$, we observe an opposite phase-dependence, which is
similar to the predicted $\pi$-junction behavior \cite{Glazman89a}. For
an intermediate coupling, $\Delta/T_K=1.8$, there is always a localized
state below the gap, which has a non-monotonic phase-dependence. In the
following, we will discuss the Josephson current through the quantum
dot.

%%%%
\paragraph{Josephson current.}
We now turn to the Josephson current through the QD in the presence of a
finite phase difference $\phi$.  Within the NRG method, the Josephson
current can be conveniently
calculated by the relation~\cite{Izumida97a}
\begin{equation}
\label{kondo-sds::eq:Is/Ic}
\frac{I_S(\phi)}{I_c^\mathrm{short}}
= -\sqrt{\frac{D\Gamma}{2\pi\Delta^2}}
\left[\sin(\phi/4)J_e + \cos(\phi/4)J_o\right]
\end{equation}
with
\begin{math}
J_\mu \equiv \sum_\sigma \left(d_\sigma^\dag f_{\mu,0,\sigma} +
  h.c.\right)
\end{math}
($\mu=e,o$).  Here
\begin{math}
I_c^\mathrm{short} \equiv {e\Delta/\hbar}
\end{math}
is the critical current of a transparent single-mode junction
\cite{Beenakker91a}.

\begin{figure}
\centering%
\includegraphics*[width=65mm]{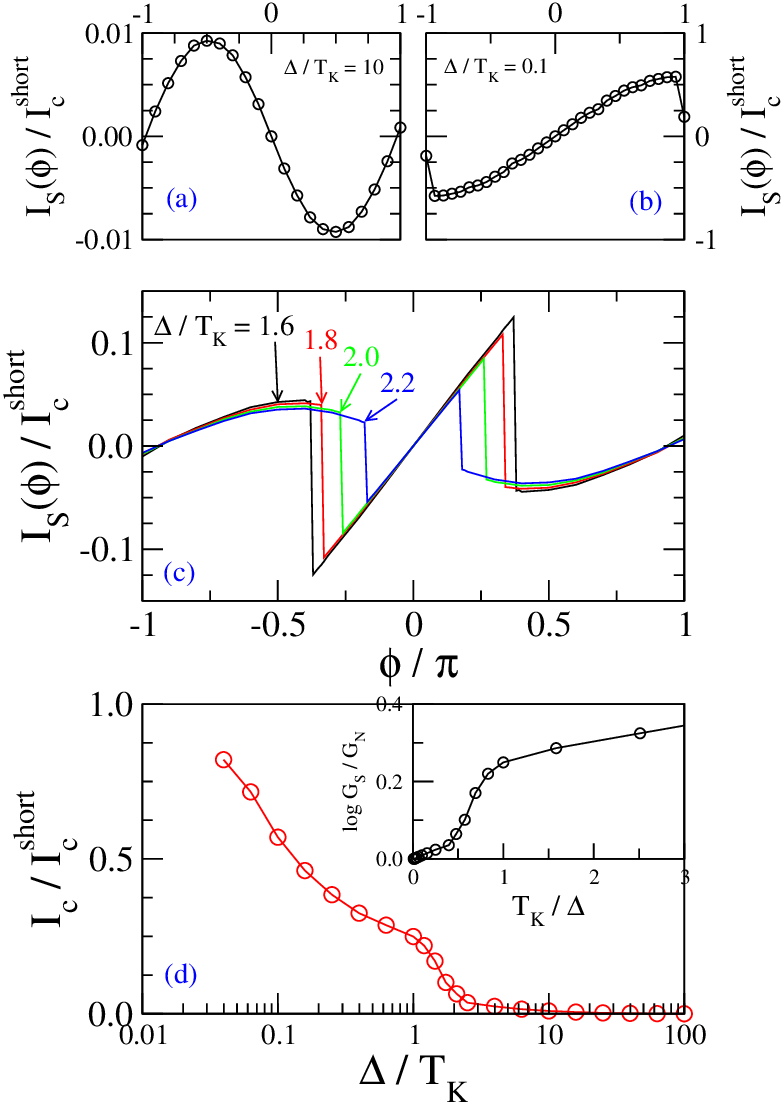}
\caption{Josephson current $I_S(\phi)$ (in units of
  $I_c^\mathrm{short}\equiv e\Delta/\hbar$) as a function of phase
  different $\phi$ (a) for $\Delta/T_K=10$ and (b) for $\Delta/T_K=0.1$.
  (c) Same curves for $\Delta/T_K=1.6$, $1.8$, $2.0$, and $2.2$ (near
  the $0$-$\pi$ junction transition point). (d) Critical current in the
  Kondo regime. We put $\eps_d=-U/2=-0.1D$ and $\Gamma=0.04D$. Inset:
  conductance resulting from the RSJ-model (see text).}
\label{kondo-sds::fig:3}
\end{figure}

Figure~\ref{kondo-sds::fig:3} shows the Josephson current as a function
of phase difference $\phi$ between the two superconducting leads for
different values of ratio $\Delta/T_K$.  In the weak coupling limit
($T_K\ll\Delta$), it is clearly seen from Fig.~\ref{kondo-sds::fig:3}
(a) that the effective Josephson coupling is negative (i.e., a
$\pi$-junction)~\cite{Shiba69a,Glazman89a,Spivak91a,Clerk00b,Rozhkov01a}.
In addition, the supercurrent-phase relation is very close to a
sinusoidal function, like typical ``tunneling
junctions''~\cite{Beenakker91a}.  We also report that the ground state
is a doublet for any phase difference $\phi$.

In the strong coupling limit ($T_K\gg\Delta$), on the other hand, the
Josephson coupling is positive~\cite{Rozhkov99a,Rozhkov00a,Clerk00b};
see Fig.~\ref{kondo-sds::fig:3} (b).  Another remarkable thing is that
the current-phase relation is highly non-sinusoidal and reminiscent of
the current-phase relation in the short junction
limit~\cite{Beenakker91a}.  Furthermore, the critical current approaches
the unitary limit $I_c^\mathrm{short}$ of ``short junctions''
\cite{Beenakker91a} as the coupling grows stronger ($\Delta/T_K\to 0$),
as shown in Fig.~\ref{kondo-sds::fig:3} (d).
These results suggest again that in the strong coupling limit the Kondo
resonance develops at the Fermi level and Cooper pairs tunnel resonantly
through it.  Naturally, the ground state turns out to be a spin singlet
for any $\phi$.  It should be stressed here that although the Kondo
effect manifests itself as a resonance channel for the Cooper-pair
tunneling, the Kondo peak of width $T_K$ in the quasi-particle
excitation spectrum is suppressed (showing a gap) below the energy scale
of order $\Delta$ ($\ll T_K$); see Fig.~\ref{kondo-sds::fig:2}.

Another interesting regime is the intermediate one ($\Delta\sim T_K$).
As demonstrated in Fig.~\ref{kondo-sds::fig:3} (c), for $\Delta\sim T_K$
the curve of $I_S(\phi)$ breaks into three distinct segments.  The
central segment resembles that of a ballistic short
junction~\cite{Beenakker91a}, while the two surrounding segments are
parts of a $\pi$-junction curve~\cite{Rozhkov99a}. Namely, the critical
value $\Delta_c(\phi)$ depends on $\phi$ with
\begin{math}
\Delta_c(\phi) > \Delta_c(\phi')
\end{math}
for $|\phi|<|\phi'|$~\cite{Choi00}; for example,
$\Delta_c(0.3\pi)\approx 1.6$ and $\Delta_c(0)\approx 2.4$.  Evidently,
the NRG results show that the ground state is a spin singlet in the
central segments ($\Delta<\Delta_c(\phi)$) and a doublet in the other
($\Delta>\Delta_c$).

%%%%
\paragraph{Experiments.}
In the experiments of Buitelaar \textit{et al.}  \cite{Buitelaar02a} the
interplay between superconductivity and Kondo physics was observed in
non-equilibrium transport (multiple Andreev
reflections)~\cite{Avishai01a,Yeyati03a}, but no supercurrent was
measured. However, the absence of a dissipationless branch in the IV is
not surprising in such (intrinsically) small junctions. Indeed thermal
or quantum fluctuations in connection with a resistive environment can
lead to a finite resistance \cite{Tinkham96a}. In
Ref.~\cite{Buitelaar02a} the ``quality factor'' $R_NC(2eI_c/\hbar
C)^{1/2}$, governing the dynamics of the corresponding
resistively-shunted junction (RSJ) model, is always smaller than 1
\cite{parameters} and the junction is therefore overdamped. In this
limit the measured resistance $G_S$ is directly related to the
current-phase relation, roughly like $G_S/G_N\sim \exp(\hbar I_c/eT)$
\cite{Ambegaokar69a}.  This enables us to relate our results of
Fig.~\ref{kondo-sds::fig:3} to the measured crossover of the conductance
as function of $T_K/\Delta$, see Fig.~4 of \cite{Buitelaar02a}.  For the
experimental temperature $T=50$mK and gap parameter $\Delta\sim1.2$K,
the calculated critical current in Fig.~\ref{kondo-sds::fig:3} (d) means
that the factor $\hbar I_c/eT$ becomes much larger than 1 in the Kondo
regime $\Delta \ll T_K$, when the transparent junction limit is reached.
Thus, the experimentally observed crossover to $G_S > G_N$ in this limit
is a manifestation of the supercurrent approaching the unitary limit
$e\Delta/\hbar$. The inset of Fig.~\ref{kondo-sds::fig:3} (d) shows the
conductance as a function of $T_K/\Delta$ and that the crossover appears
for $T_K\approx 0.5 \Delta$, which is in quite good agreement with the
experimental result of Ref.~\cite{Buitelaar02a}.

\paragraph{Conclusion.}
We have studied the Josephson current and the proximity effect on the QD
coupled to two SCs in a whole range of coupling.  Our results exhibit a
transition from the weak to the strong coupling limit, which occurs when
$\Delta\sim{}T_K$.  In the weak coupling limit, superconductivity
dominates the Kondo physics, and the tunneling of Cooper pairs can be
treated perturbatively.  The system is a $\pi$-junction, the pairing
correlation on the QD is negative, and the ground state is a spin
doublet.  In the strong coupling limit, the Kondo effect becomes
important and manifests itself as a resonance channel for the
Cooper-pair tunneling.  This leads to a positive Josephson coupling
($0$-junction) and positive pairing correlation on the QD.  Here the
Kondo effect in the presence of superconductivity is distinguished from
the usual one with normal leads in that the Kondo peak in the
quasi-particle excitation spectrum is suppressed completely (exhibiting
a gap) for energies below the superconducting gap.

\paragraph{Acknowledgements.}
We thank C.~Sch\"onenberger for remarks on Ref.~\cite{Buitelaar02a} and
acknowledge discussions with C. Bruder, M.~Gr\"aber, and T.~Kontos.
This work was supported by the SKORE-A, the eSSC at Postech, the NCCR
Nanoscience, and the Swiss NSF.

%%%% References
%% \bibliographystyle{physrev}
%% \bibliography{aliases,cond-mat,staphy,physics,qubits,choims,kondo-sds}%

\begin{thebibliography}{10}
\bibitem{endnote:1} With $\phi=0$, we map the model in
  Eqs.~(\ref{kondo-sds::eq:HBCS})--(\ref{kondo-sds::eq:HV}) to an
  effective model with only a single lead, for which the NRG calculation
  is far more efficient than for the two-lead problem.
\bibitem{parameters} The capacitance of the junction is of the order of
  $C=100$aF. Unfortunately the environmental resistance is not known,
  but is probably smaller than $h/e^2$ since the contact pads are
  superconducting.
\end{thebibliography}

\end{document}